\pdfoutput=1

\documentclass[review, 12pt, a4paper]{elsarticle}
\usepackage{hyperref}
\usepackage{lineno}
\modulolinenumbers[5]

\journal{Journal of \LaTeX\ Templates}

\usepackage{epstopdf}

\usepackage{numcompress}

\biboptions{sort&compress}

\begin{document}

\begin{frontmatter}


\title{Magnetoelectronic and optical properties of nonuniform graphene nanoribbons}


\author[address_NCKU,address_CMNST]{Hsien-Ching Chung\corref{mycorrespondingauthor}}
\ead{hsienching.chung@gmail.com}

\author[address_NCKU]{Yu-Tsung Lin}

\author[address_NCKU]{Shih-Yang Lin}

\author[address_TUT]{Ching-Hong Ho}

\author[address_TUT]{Cheng-Peng Chang}

\author[address_NCKU]{Ming-Fa Lin\corref{mycorrespondingauthor}}
\cortext[mycorrespondingauthor]{Corresponding author}
\ead{mflin@mail.ncku.edu.tw}

\address[address_NCKU]{Department of Physics, National Cheng Kung University, Tainan 70101, Taiwan}
\address[address_CMNST]{Center for Micro/Nano Science and Technology (CMNST), National Cheng Kung University, Tainan 70101, Taiwan}
\address[address_TUT]{Center for General Education, Tainan University of Technology, Tainan 701, Taiwan}

\begin{abstract}


The electronic and optical properties of nonuniform bilayer graphene nanoribbons are worth investigating as they exhibit rich magnetic quantization.
Based on our numerical results, their electronic and optical properties strongly depend on the competition between magnetic quantization, lateral confinement, and stacking configuration.
The results of our calculations lead to four categories of magneto-electronic energy spectra, namely monolayer-like, bilayer-like, coexistent, and irregular quasi-Landau-level like.
Various types of spectra described in this paper are mainly characterized by unusual spatial distributions of wave functions in the system under study.
In our paper, we demonstrate that these unusual quantized modes lead to the appearance of such diverse magneto-optical spectra.
Moreover, the investigation of the density of states in our model leads to the appearance of many prominent symmetric and weakly asymmetric peaks.
The almost well-behaved quasi-Landau levels exhibit high-intensity peaks with specific selection rules, and the distorted energy subbands present numerous low-intensity peaks without any selection rules.

\end{abstract}


\end{frontmatter}


\section{Introduction}
\label{sec:Introduction}

\begin{figure}
\begin{center}
  \includegraphics[width=\linewidth, keepaspectratio]{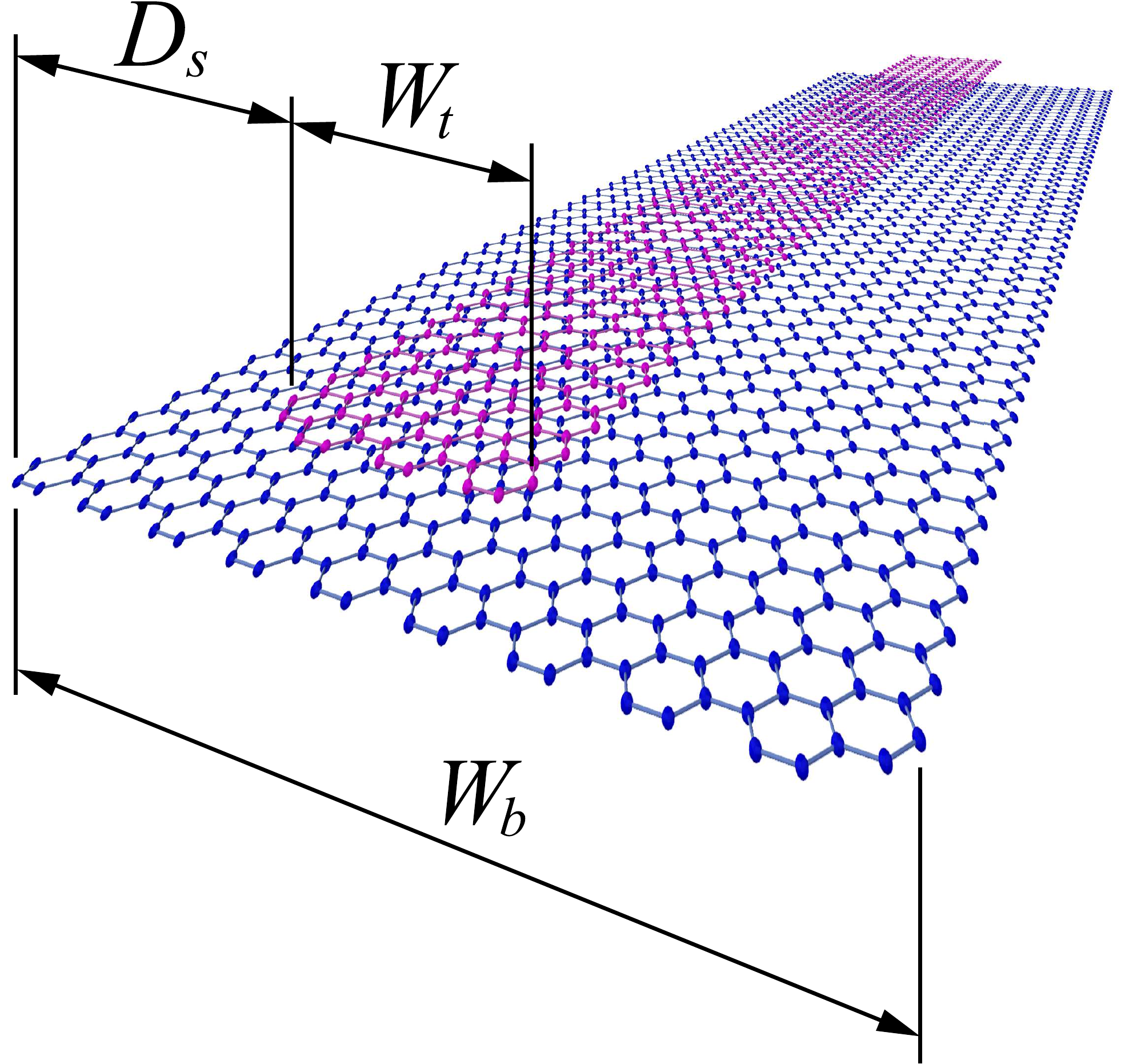}\\
  \caption[]{
  Nonuniform bilayer GNR with a narrow top layer (red strip) of width $W_t$ and a wide bottom layer (blue strip) of width $W_b$; $D_s$ is the distance between their edges. 
  }
  \label{fig:Conceptual_Illustration_Of_NonUniGNR}
\end{center}
\end{figure}

Graphene, a one-atom-thick hexagonal lattice comprising carbon atoms, was discovered in 2004~\cite{Science306(2004)666K.S.Novoselov}.
Its superior electronic~\cite{Science312(2006)1191C.Berger, Phys.Rev.Lett.100(2008)016602S.V.Morozov}, magnetic~\cite{Phys.Rev.B77(2008)085426Y.H.Lai}, optical~\cite{Science320(2008)1308R.R.Nair}, thermal~\cite{Phys.Rev.Lett.100(2008)016602S.V.Morozov, NanoLett.8(2008)902A.A.Balandin}, mechanical~\cite{Science321(2008)385C.Lee}, and transport~\cite{Nature438(2005)197K.S.Novoselov, Science315(2007)1379K.S.Novoselov} properties are of interest because of the fundamental science included and its technical applications~\cite{Appl.Phys.Lett.99(2011)163102Y.H.Su, Phys.Rev.Lett.113(2014)266801L.E.F.FoaTorres, ACSSustainableChem.Eng.3(2015)1965Y.H.Su, ACSNano9(2015)8967H.C.Wu}.
However, its gapless feature results in a low on/off ratio ($\approx 5$) in graphene-based field-effect transistors (FETs) and is an impediment to the development of graphene nanoelectronics~\cite{NanoLett.10(2010)715F.Xia}.
One of the most promising approaches to controlling the electronic and optical properties is to fabricate quasi-one-dimensional (quasi-1D) graphene strips, referred to as graphene nanoribbons (GNRs).
Electrical transport experiments have shown that the lateral-confinement-induced gap in GNRs facilitates the fabrication of FETs with high on/off ratios, approximately $10^7$ at room temperature~\cite{Science319(2008)1229X.L.Li}.
Four major state-of-the-art fabrication methods, including both top-down and bottom-up methods, have been proposed to achieve large-scale production of GNRs: cutting of graphene~\cite{Phys.Rev.Lett.98(2007)206805M.Y.Han, Science319(2008)1229X.L.Li}, unzipping of carbon nanotube (CNT)~\cite{Nature458(2009)872D.V.Kosynkin, Carbon48(2010)2596F.Cataldo}, chemical vapor deposition (CVD)~\cite{NanoLett.8(2008)2773J.Campos-Delgado, Nat.Nanotechnol.5(2010)727M.Sprinkle}, and piecewise linking of molecular precursor monomers~\cite{Nature466(2010)470J.M.Cai, Appl.Phys.Lett.105(2014)023101Y.Zhang}.
Researchers are constantly in quest of other methods to precisely control the nanoscale width and realize a perfect edge structure.
The first three of the aforementioned four methods can be used to successfully fabricate few-layer GNRs and also nonuniform GNRs.
In this study, we investigated the rich magnetic quantization of the electronic and optical properties of nonuniform bilayer GNRs.
These GNRs comprise two GNRs with different widths (Fig.~\ref{fig:Conceptual_Illustration_Of_NonUniGNR}) and can be fabricated through mechanical exfoliation~\cite{Phys.Rev.B79(2009)235415C.P.Puls, Phys.Rev.B88(2013)125410J.Tian}, cutting of graphene~\cite{Nat.Nanotechnol.3(2008)397L.Tapaszto, J.Am.Chem.Soc.132(2010)10034S.Fujii}, chemical unzipping of CNT~\cite{Nature458(2009)872D.V.Kosynkin, ACSNano5(2011)968D.V.Kosynkin}, and CVD~\cite{NanoLett.8(2008)2773J.Campos-Delgado, J.Am.Chem.Soc.131(2009)11147D.C.Wei}.


Changes in the edge structure, ribbon width, and stacking configuration can drastically alter the electronic properties of GNRs.
There are two achiral edge structures: zigzag and armchair.
The energy gaps of armchair GNRs (AGNRs) have been predicted to be inversely proportional to the GNR width~\cite{NanoLett.6(2006)2748V.Barone, Adv.Phys.59(2010)261D.S.L.Abergel}.
By contrast, zigzag GNRs (ZGNRs) show dispersionless partial flat subbands at the Fermi energy ($E_F = 0$), arising from the edge-localized states~\cite{J.Phys.Soc.Jpn.65(1996)1920M.Fujita, Phys.Rev.B73(2006)235411L.Brey}.
The width-dependent energy gap has been confirmed by conductance~\cite{Phys.Rev.Lett.98(2007)206805M.Y.Han, Science319(2008)1229X.L.Li} and tunneling current measurements~\cite{Nat.Nanotechnol.3(2008)397L.Tapaszto}, and the existence of the edge-localized states has been verified using scanning tunneling microscopy (STM)~\cite{Phys.Rev.B71(2005)193406Y.Kobayashi, Nat.Commun.5(2014)4311Y.Y.Li} and angle-resolved photoemission spectroscopy~\cite{Phys.Rev.B73(2006)045124K.Sugawara}.
There are two typical stackings in GNRs, namely AB and AA stackings.
In the AA stacking, all carbon atoms of the hexagonal rings are nearest-neighbors (i.e., positioned on top of each other), whereas in the AB stacking, only half of the atoms are nearest-neighbors, with the remaining half of the atoms being located above and below the empty centers of the hexagonal rings.
Bilayer GNRs show two groups of energy subbands, with each group containing both conduction and valence subbands.
The onset energies of the subband groups are closely related to the stacking configuration.
For the high symmetry AA stacking, both groups are initiated at an energy approximately equal to the vertical  nearest-neighbor interlayer interaction, and there is a strong overlap between the valence subbands of the first group and the conduction subbands of the second group.
By contrast, for the AB stacking, one group is at the Fermi energy and the other group is displaced by an energy equal to the nearest-neighbor interlayer interaction.
Moreover, there is only a weak overlap between the conduction and the valence subbands of the first group~\cite{Phys.Chem.Chem.Phys.18(2016)7573H.C.Chung}.

A uniform perpendicular magnetic field can bunch neighboring electronic states and thereby form highly degenerate Landau levels with quantized cyclotron radii. 
Magnetic quantization is strongly suppressed by the lateral confinement.
The competition between magnetic quantization and lateral confinement leads to a variety of magnetoelectronic structures including partly dispersionless quasi-Landau levels (QLLs), partial flat subbands, and 1D parabolic subbands~\cite{Phys.Rev.B59(1999)8271K.Wakabayashi, Phys.Rev.B91(2015)155409B.Ostahie}.
The energy dispersions associated with these subbands lead to two types of peaks in the density of states (DOS), symmetric and asymmetric peaks~\cite{Nanotechnol.18(2007)495401Y.C.Huang, PhysicaE42(2010)711H.C.Chung, J.Phys.Soc.Jpn.80(2011)044602H.C.Chung}.
QLLs can be formed for sufficiently wide nanoribbons, and each Landau wave function has a localized symmetric/antisymmetric spatial distribution with a specific number of zero points (quantum number $n$).
The magneto-optical spectra exhibit numerous symmetric and asymmetric absorption peaks.
The symmetric absorption peaks originate from inter-QLL transitions and obey the edge-independent magneto-optical selection rule $| \Delta n | = 1$~\cite{Nanotechnol.18(2007)495401Y.C.Huang}.
By contrast, the asymmetric peaks originate from optical transitions among parabolic subbands and satisfy the edge-dependent selection rules~\cite{Phys.Rev.B76(2007)045418H.Hsu, J.Phys.Soc.Jpn.69(2000)3529M.F.Lin, Opt.Express19(2011)23350H.C.Chung}.
When the number of layers is doubled, bilayer GNRs have two groups of QLLs~\cite{PhysicaE42(2010)711H.C.Chung, Phys.Chem.Chem.Phys.15(2013)868H.C.Chung, Philos.Mag.94(2014)1859H.C.Chung}, and their onset energies are the same as the zero-field energies.
Furthermore, only the intragroup optical transitions occur in the AA stacking, whereas intergroup transitions are also occur in the AB stacking.
This difference is derived from the fact that the wave functions of the AA stacking are symmetric or antisymmetric superpositions of the tight-binding functions on different layers.

The magnetoelectronic and optical properties of nonuniform bilayer GNRs are directly dependent on the geometric structures, such as the widths of the top and bottom layers ($W_t$ and $W_b$), the relative distance between their edges ($D_s$), and inner boundaries of the top narrow ribbon (Fig.~\ref{fig:Conceptual_Illustration_Of_NonUniGNR}).
Four categories of magnetoelectronic energy spectra are reflected in the DOS, and they involve monolayer-like, bilayer-like, coexistent, and irregular QLLs.
The various categories of QLL spectra show distinct optical properties.
In particular, the coexistent QLLs can exist only when the widths of the monolayer and bilayer regions are considerably greater than the magnetic length.
The QLL wave functions are well behaved, and they give rise to many high-intensity absorption peaks with two specific optical selection rules.
In GNRs with irregular QLLs, the two subband groups of different layers vary considerably, and the wave functions are seriously distorted and piecewise continuous.
Consequently, many low-intensity absorption peaks without any selection rules exist.
These feature-rich electronic and optical properties can be experimentally verified, and the nonuniform GNR provides an alternative for understanding the optical properties of other related structures, such as partially overlapping GNRs~\cite{Science336(2012)1143A.W.Tsen, Carbon75(2014)411M.Berahman, Carbon80(2014)513R.Rao}, GNR-graphene superlattices~\cite{J.Phys.Chem.C117(2013)7326J.H.Wong}, and van der Waals heterostructures~\cite{Nature499(2013)419A.K.Geim, Nanoscale7(2015)13393Z.Zheng, Phys.Rev.B93(2016)075438L.E.F.FoaTorres}.

\section{Tight-binding model}

\begin{figure*}
\begin{center}
  \includegraphics[width=\linewidth]{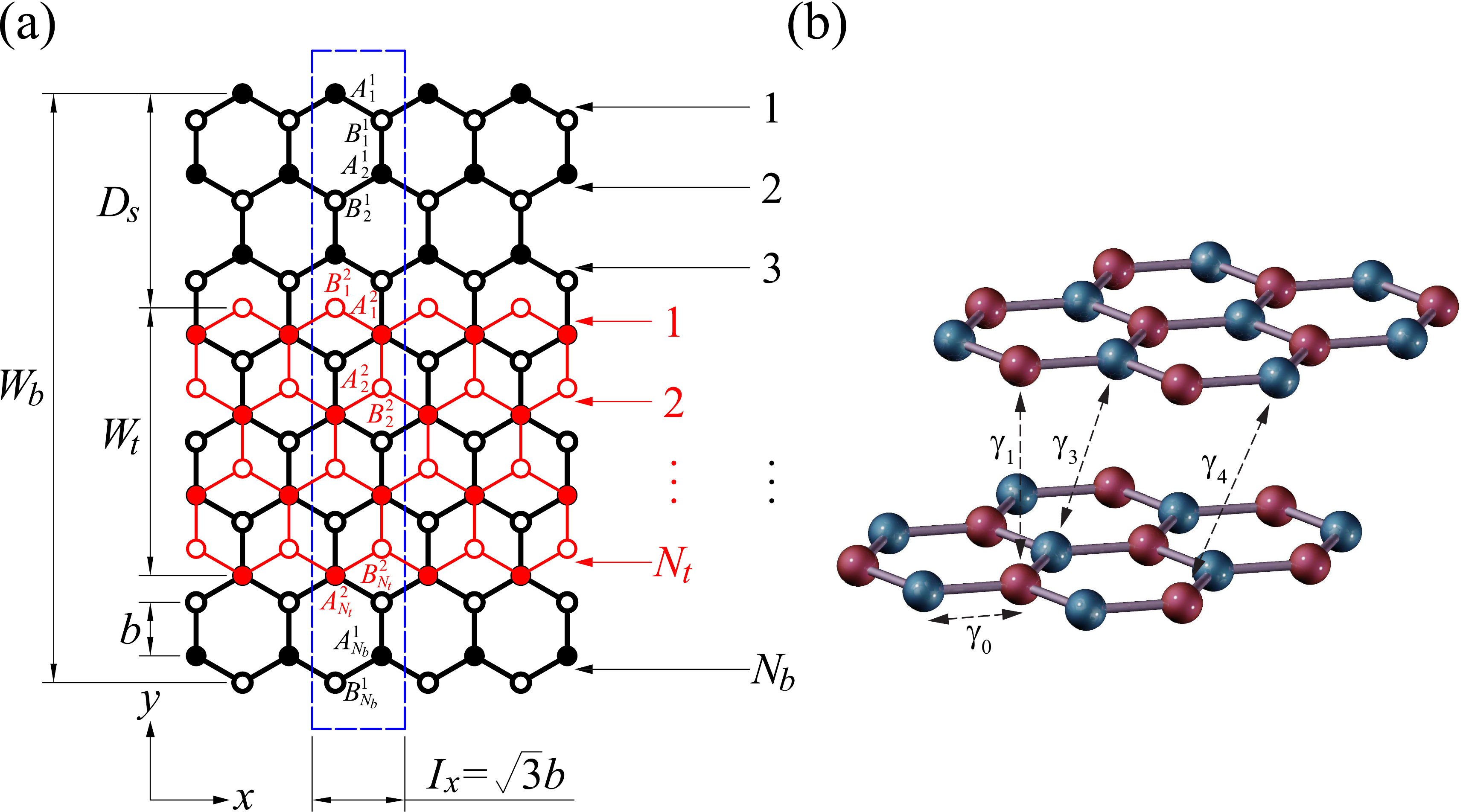}\\
  \caption[]{
  (a) Geometric structure of a nonuniform bilayer AB-stacked ZGNR.
  (b) Intralayer and interlayer atomic interactions.
  }
  \label{fig:Geometric_Structure_of_NonUniGNR}
\end{center}
\end{figure*}

Zigzag and armchair nanoribbons feature similar magneto-optical transitions, thus for the sake of brevity, we focus on the study of nonuniform bilayer AB-stacked ZGNRs to investigate the edge-independent state energy of their QLLs and magneto-optical selection rule.
The geometric structure is composed of two ZGNRs of different widths (Fig.~\ref{fig:Geometric_Structure_of_NonUniGNR}(a)).
The wide bottom and narrow top layers are represented by the heavy black and light red hexagonal lattices, respectively.
The primitive unit cell enclosed by the dashed rectangle has a periodic length given by $I_x = \sqrt{3}b$, where $b=1.42$ \AA~is the C--C bond length.
The first Brillouin zone is in the region $-\pi /I_x < k_x \leq \pi /I_x$.
The carbon atoms on the \emph{m}th zigzag line of the bottom (top) layer are denoted by $A^{1(2)}_m$'s (filled circles) or $B^{1(2)}_m$'s (open circles).
The superscripts 1 and 2 indicate the top and bottom layers, respectively.
There are $2N_b + 2N_t$ carbon atoms in a unit cell, where $N_b$ ($N_t$) is the number of zigzag lines of the bottom (top) layer.
The top and bottom layers have widths of $W_t = b(3N_t/2 - 1)$ and $W_b = b(3N_b/2 - 1)$, respectively, and are vertically shifted by a distance of $c = 3.35$~{\AA}.
The layers conform to AB stacking; in other words, the two hexagonal lattices are displaced by a horizontal distance $b$ from each other as viewed from the top.
The first zigzag line of the top layer is on the ($d_s+1$)th zigzag line of the bottom layer, and the distance between the edges of the top and bottom layers is $D_s = | b(3d_s-1)/2 |$.


The low-energy electronic properties are dominated by the tight-binding functions of $2p_z$ orbitals.
On the basis of the $2N_b + 2N_t$ tight-binding functions, the nearest-neighbor Hamiltonian can be expressed as
\begin{eqnarray}
\nonumber
\mathcal{H} &=& -\gamma_0 \displaystyle\sum_{m,n}
\bigg( |\Phi^{A^1}_m\rangle \langle \Phi^{B^1}_n| +
|\Phi^{A^2}_m\rangle \langle \Phi^{B^2}_n| \bigg) \\
&&-\displaystyle\sum_i \gamma_i \sum_{m,n}
|\Phi^{A^1}_m\rangle \langle \Phi^{B^2}_n|
 + \mathrm{H.c.}.
\label{eq:TightBindingHamiltonianOfNonUniGNRs}
\end{eqnarray}
The Bloch wave function is given by
\begin{eqnarray}
\nonumber
| \Psi^{c,v} \rangle & = & \displaystyle\sum_m^{N_b}
\bigg( A^1_m |\Phi^{A^1}_m\rangle + B^1_m |\Phi^{B^1}_m\rangle \bigg) \\
& + & \displaystyle\sum_n^{N_t}
\bigg( A^2_n |\Phi^{A^2}_n\rangle + B^2_n |\Phi^{B^2}_n\rangle \bigg) ,
\label{eq:WFOfNonUniGNRs}
\end{eqnarray}
where $|\Phi^{A^{1(2)}}_m\rangle$ ($|\Phi^{B^{1(2)}}_n\rangle$) represents the tight-binding functions associated with the $2p_z$ orbitals on the $m$th ($n$th) site of the first-layer (second-layer) sublattice $A^{1(2)}$ ($B^{1(2)}$).
The first term describes only the intralayer atomic interactions between the nearest neighbors in each layer, and the second term pertains to the interlayer atomic interactions.
According to the Slonczewski--Weiss--McClure model~\cite{Phys.Rev.71(1947)622P.R.Wallace}, the atomic interactions in $2p_z$ orbitals for the AB stacking are the intralayer atomic interactions between the nearest carbon atoms ($\gamma_0 = 2.598$ eV), and interlayer atomic interactions between $A^1$ and $A^2$ atoms ($\gamma_1 = 0.364$ eV), $B^1$ and $B^2$  atoms ($\gamma_3 = 0.319$ eV), and $A^{1}$ and $B^{2}$ ($A^{2}$ and $B^{1}$) atoms ($\gamma_4 = 0.177$ eV) (Fig.~\ref{fig:Geometric_Structure_of_NonUniGNR}(b)); there is also a chemical shift ($\gamma_6 = -0.026$ eV)~\cite{Phys.Rev.B43(1991)4579J.C.Charlier}.
$A^1_m$, $B^1_m$, $A^2_m$, and $B^2_m$ in Eq. (\ref{eq:WFOfNonUniGNRs}) are the amplitudes of the subenvelope functions on the four distinct sublattices.

When a uniform static magnetic field given by $\mathbf{B}=B_0 \hat{z}$ is applied perpendicular to the ribbon plane, the effective Hamiltonian can be interpreted as the Peierls substitution Hamiltonian~\cite{Z.Phys.80(1933)763R.Peierls}.
Each Hamiltonian matrix element becomes the product of the zero-field element and the extra phase factor $\exp (i2\pi \theta _{mn})$, where $\theta_{mn}$ is expressed as
$(1 / \phi_{0}) \int_{m}^{n}\mathbf{A}\cdot d\mathbf{l}$
and is defined as a line integral of the vector potential $\mathbf{A}$ from the $m$th site to the $n$th site.
Here, $\mathbf{A}$ is chosen as $(-B_0 y,0,0)$ to preserve the translational invariance in the $x$-direction under the Landau gauge, and $\phi_0=h/e$ is the magnetic flux quantum.
The diagonalization of the Hamiltonian matrix $\mathcal{H}$ yields the energy spectrum $E^{c,v}(k_x)$ and wave functions $| \Psi^{c,v} \rangle$, where the superscripts $c$ and $v$ denote the conduction and valence subbands, respectively.


\section{Magnetoelectronic properties}

\begin{figure}
\begin{center}
  \includegraphics[width=0.98\linewidth, keepaspectratio]{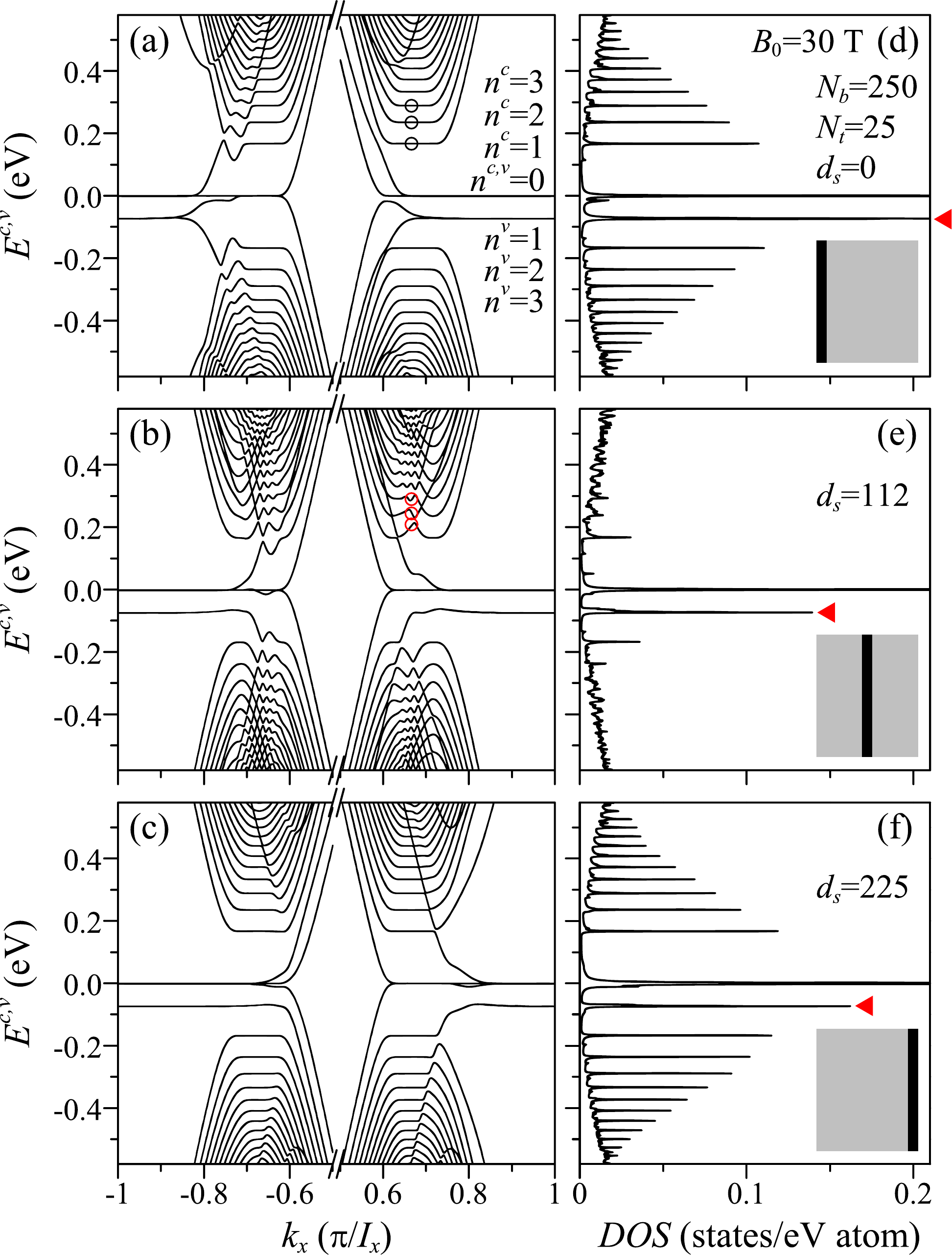}\\
  \caption[]{
  Low-energy magnetoelectronic structures of nonuniform bilayer AB-stacked ZGNRs with $N_b = 250$ and $N_t = 25$ at (a) $d_s =0$, (b) $d_s = 112$, and (c) $d_s = 225$ for $B_0 = 30$ T.
  The top view of each configuration is depicted in the gray zones.
  The circles mark the locations of the electronic states on the $n^c = 1$--$3$ QLLs at $k_x = 2/3$.
  The DOSs are presented in (d), (e), and (f).
  The red triangles indicate the peaks at $\omega = -75$ meV.
  }
  \label{fig:BS_DOS_Po_shift_NonUniZGNR}
\end{center}
\end{figure}

The electronic properties of nonuniform bilayer ZGNRs are very sensitive to changes in the relative position between the two layers, ribbon width, interlayer atomic interactions, and magnetic field.
There are four categories of magnetoelectronic energy spectra, namely monolayer-like, bilayer-like, coexistent, and irregular QLL spectra.
First, when the top layer is insufficiently wide for magnetic quantization and the bottom layer is sufficiently wide for magnetic quantization, the energy spectrum can be regarded as a mixture of two monolayer-like spectra.
One monolayer-like spectrum consists of monolayer-like QLLs contributed by the bottom layer, whose width is greater than the magnetic length ($l_B = \sqrt{\hbar /eB_0}$).
The other monolayer-like spectrum comprises the parabolic subbands arising from the top layer, which has a width lesser than $l_B$.
The relative position between the two layers plays a crucial role in the mixing of these electronic energy spectra.
As the top layer is at the edge of the bottom layer ($d_s = 0$), the Landau orbits near the center of the bottom layer are almost unaffected (Fig.~\ref{fig:BS_DOS_Po_shift_NonUniZGNR}(a)).
The QLLs are formed around $k_x = 2/3$, and their energies are roughly proportional to $\sqrt{n^{c,v}B_0}$, where $n^{c,v} = 0, 1, 2,...$ is the subband index labeled from the Fermi energy.
QLLs with a larger-$n^{c,v}$ have wider Landau orbits and a smaller $k_x$ range in the dispersionless energy spectrum.
Because of the destruction of inversion symmetry, the energy spectrum is no longer symmetric about the $k_x = 0$ axis ($E^{c,v}(k_x) \neq E^{c,v}(-k_x)$).
Apparently, the parabolic subbands at $-k_x$ mix with the left-hand side of the QLLs, instead of the right-hand side of QLLs.
Furthermore, two pairs of degenerate partial flat subbands are revealed near $E_F = 0$ with an energy spacing $\Delta U_I$ of $75$ meV, and they are one of the main features of bilayer ZGNRs.
When the top layer gradually moves to the center, the parabolic subbands shift to the formation center of QLLs ($k_x = \pm 2/3$, where parabolic subbands become QLLs as the magnetic field grows) and subband mixings are considerably enhanced.
In particular, for the case of a center-positioned top layer ($d_s = 112$), the dispersionless QLLs are transformed to irregular oscillatory QLLs with many extra band-edge states (Fig.~\ref{fig:BS_DOS_Po_shift_NonUniZGNR}(b)), indicating that the Landau orbits are strongly suppressed (this is discussed later).
The dispersionless QLLs can be recovered from the irregular oscillatory QLLs, as the top layer moves to the right edge ($d_s = 225$ in Fig.~\ref{fig:BS_DOS_Po_shift_NonUniZGNR}(c)).
Furthermore, the parabolic subbands are shifted to the right side of the QLLs.
Additionally, the energy spectra for the cases of the left-edge- and right-edge-aligned top layer (Figs.~\ref{fig:BS_DOS_Po_shift_NonUniZGNR}(a) and~\ref{fig:BS_DOS_Po_shift_NonUniZGNR}(c)) are not symmetric about the $k_x = 0$ axis because of the asymmetry in the geometric structures.

\begin{figure}
\begin{center}
  \includegraphics[width=0.95\linewidth, keepaspectratio]{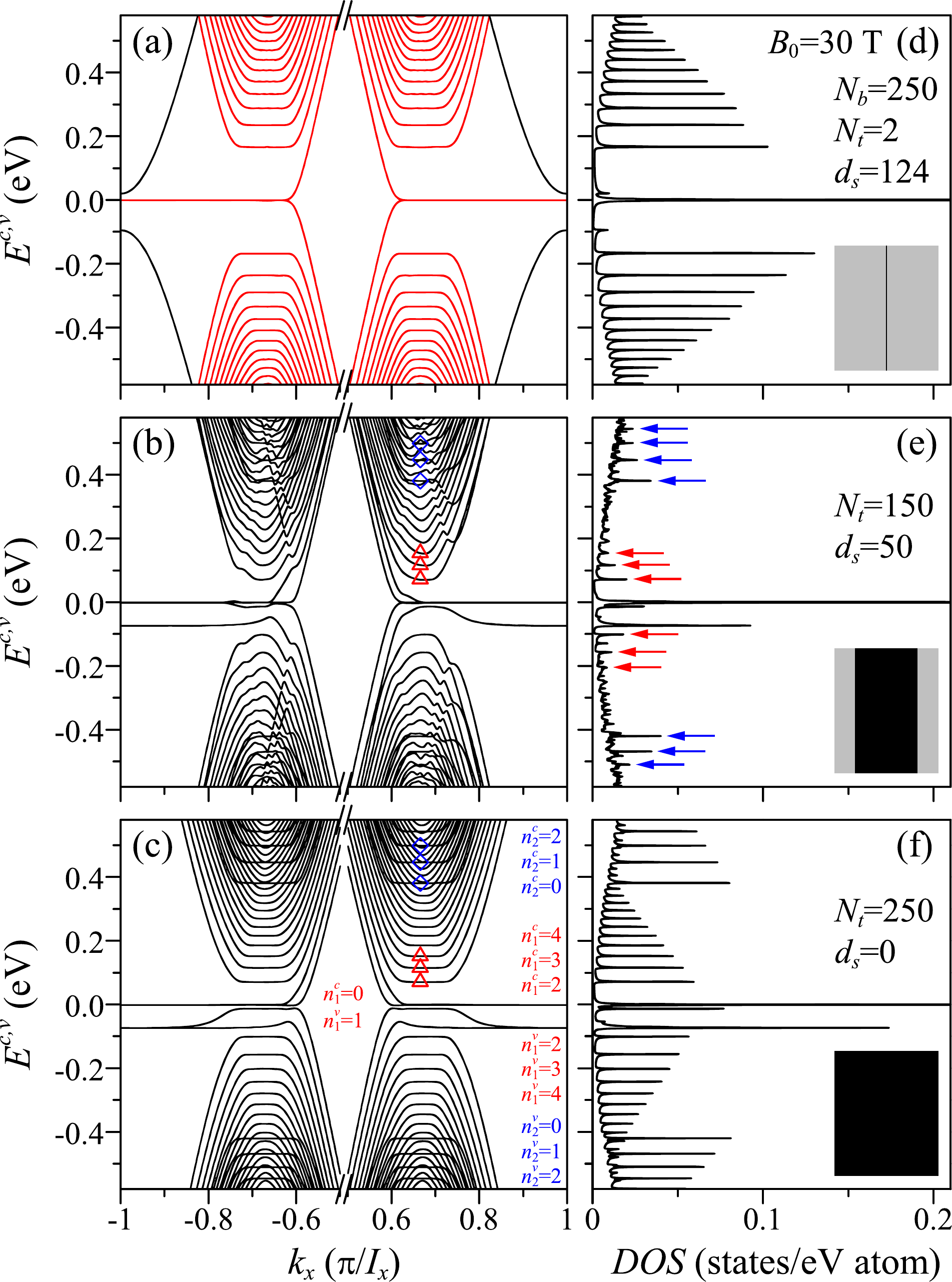}\\
  \caption[]{
  Magnetoelectronic structures of the nonuniform ZGNRs with $N_b = 250$ for (a) $N_t = 2$, (b) $N_t = 150$, and (c) $N_t = 250$ at the ribbon center.
  The red triangles and blue diamonds mark the locations of the the electronic states on the first- and second-group QLLs at $k_x = 2/3$, respectively.
  (d)--(f) The corresponding DOSs.
  The red and blue arrows indicate the first- and second-group QLL peaks, respectively.
  }
  \label{fig:BS_DOS_Width_var_NonUniZGNR}
\end{center}
\end{figure}

The QLL spectrum can be transformed from a monolayer-like spectrum to a bilayer-like spectrum, depending on the top layer width.
For an extremely narrow top layer ($N_t = 2$, $W_t < l_B$, and $W_b > l_B$), the main magnetoelectronic energy spectrum is the energy dispersions of the bottom layer, including the monolayer-like QLLs and the partial flat subbands at $E_F = 0$ (red curves in Fig.~\ref{fig:BS_DOS_Width_var_NonUniZGNR}(a)).
The top layer contributes only two parabolic subbands with band-edge states at the zone boundaries (black curves).
With an increase in the top layer width ($N_t = 25$, $W_t \sim l_B$, and $W_b > l_B$ in Fig.~\ref{fig:BS_DOS_Po_shift_NonUniZGNR}(b)), the monolayer-like QLL spectrum is highly distorted, exhibiting serious subband mixing, because of the interlayer atomic interactions.
Later, the second group of QLLs starts at the energies of the nearest-neighbor interlayer interactions ($\pm \gamma_1$) when $W_t > l_B$ ($N_t = 150$ in Fig.~\ref{fig:BS_DOS_Width_var_NonUniZGNR}(b)).
The monolayer-like QLL spectrum evolves into a bilayer-like QLL spectrum.
The energy spacings between QLLs are reduced, and the $\sqrt{B_0}$ dependence of QLL energies no longer holds.
It is noteworthy that the two groups of QLLs are mixed with extra parabolic subbands.
The parabolic subbands are due to the nonoverlapping regions of the bottom layer, whose width is insufficiently for magnetic quantization.
In another extreme case (Fig.~\ref{fig:BS_DOS_Width_var_NonUniZGNR}(c)), the top layer becomes as wide as the bottom layer ($N_t = 250$ and $W_t = W_b > l_B$), and consequently the entire bilayer QLL spectrum is obtained; in Fig.~\ref{fig:BS_DOS_Width_var_NonUniZGNR}(c), the QLLs of the first and second group are indexed with $n_1^{c,v}$ and $n_2^{c,v}$, respectively.

\begin{figure*}
\begin{center}
  \includegraphics[width=\linewidth, keepaspectratio]{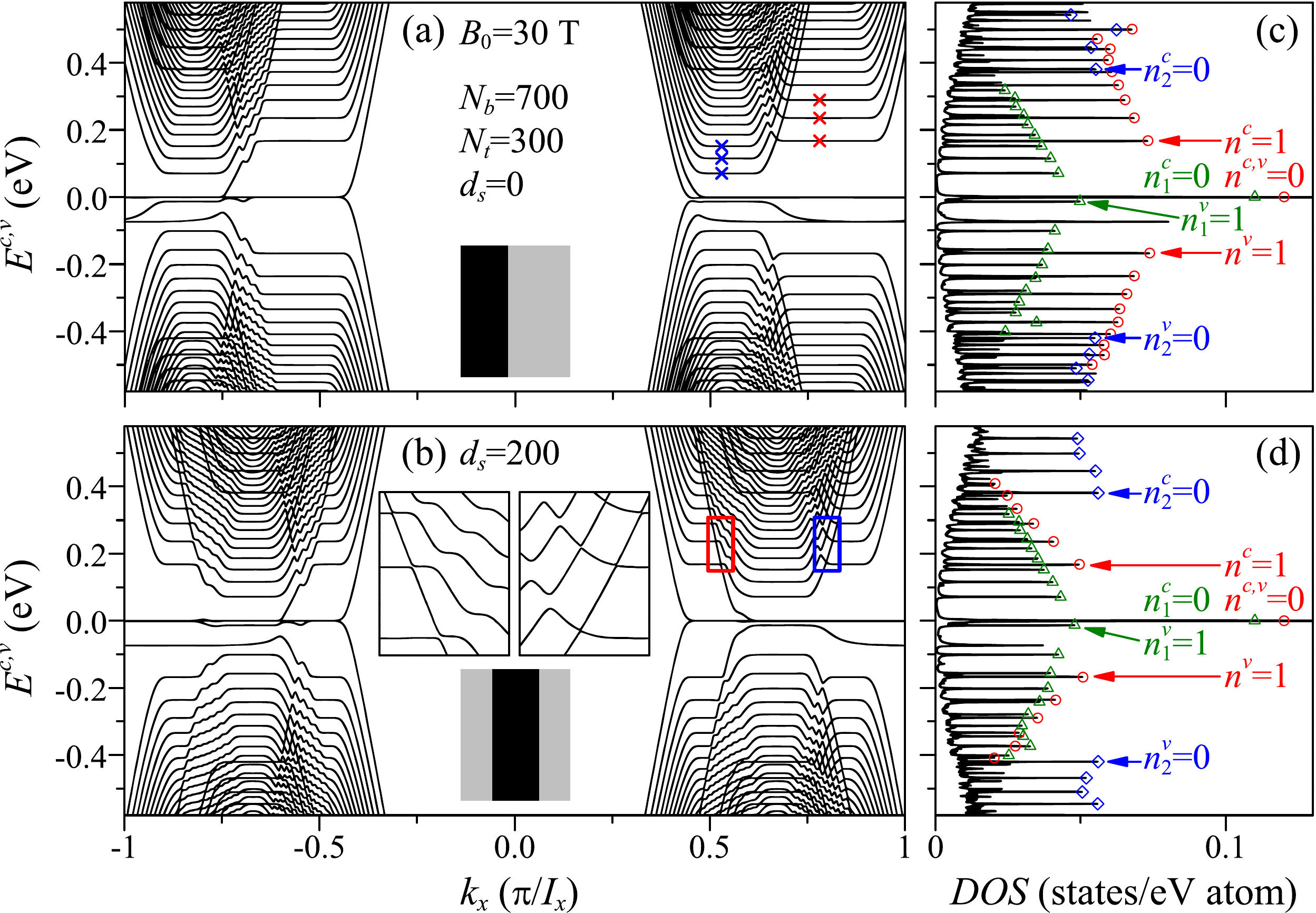}\\
  \caption[]{
  Coexistence of monolayer- and bilayer-like QLL energy spectra for nonuniform ZGNRs with $N_b = 700$ and $N_t = 300$ at (a) $d_s = 0$ and (b) $d_s = 200$.
  Insets show magnified views of the intermediate QLLs within the red and blue boxes.
  The red and blue crosses mark the locations of the electronic states at $k_x = k_1$ and $k_2$, respectively.
  The corresponding DOSs are shown in (c) and (d), where the red circles, green triangles, and blue diamonds indicate the first few monolayer-like QLL peaks,  bilayer-like QLL peaks of the first group, and bilayer-like QLL peaks of the second group, respectively.
  }
  \label{fig:BS_DOS_Coexistence_NonUniZGNR}
\end{center}
\end{figure*}

The monolayer- and bilayer-like QLL spectra can simultaneously exist because the overlapping and nonoverlapping regions are sufficiently wide for quantizing the electronic states.
Furthermore, bilayer- and monolayer-like QLLs at positive (negative) wave vectors are obtained at $k_x < 2/3$ and $k_x > 2/3$ ($k_x < -2/3$ and $k_x > -2/3$), respectively (Fig.~\ref{fig:BS_DOS_Coexistence_NonUniZGNR}(a)).
An extra group of intermediate QLLs is formed, and it shows subband mixing of the monolayer- and bilayer-like QLLs because of the interface between the regions with/without interlayer interactions.
The spatial distributions of intermediate states exhibit wave functions at the interface, and they are discussed later.
The coexistent QLL spectrum is also sensitive to the top layer position.
In the case of the center-positioned top layer (Fig.~\ref{fig:BS_DOS_Coexistence_NonUniZGNR}(b)), there are two groups of monolayer-like QLLs, one group of bilayer-like QLLs, and two extra groups of intermediate QLLs, reflecting the special geometric structure of the nonuniform bilayer ZGNR with two interfaces.

The DOS, defined as $2\sum_{k_x, n^{c,v}} \delta [\omega - E^{c,v}(k_x, n^{c,v})]$ (the numeral $2$ indicates the spin), directly reflects the main features of the energy spectra~\cite{Sci.Rep.5(2015)9423P.Y.Lo}.
Two types of peaks are presented, including symmetric and asymmetric peaks.
The symmetric peaks are delta-function-like peaks with two-side-divergent structure at the peak frequency $\omega$ (so that the peak is symmetric about the axis of $\omega$).
They are the prominent peaks and originate from partial flat subbands and dispersionless QLLs.
The asymmetric peaks with one-side-divergent structure are the secondary peaks, and they originate from the band-edge states of parabolic and irregular oscillatory subbands.
The edge-aligned narrow top layer partially distorts the QLLs, and therefore, the monolayer-like DOS is almost unaffected (Figs.~\ref{fig:BS_DOS_Po_shift_NonUniZGNR}(d) and~\ref{fig:BS_DOS_Po_shift_NonUniZGNR}(f)).
The highest symmetric peak at $E_F = 0$ originates mainly from the partial flat subbands, and the QLL peak intensity, proportional to the QLL spread range in the momentum space, gradually decreases with an increase in $n^{c,v}$.
The high-intensity QLL peaks are surrounded by asymmetric subpeaks, which are attributed to the extra band-edge states of the mixed subbands.
Notably, the symmetric peak of high intensity at $\omega = -75$ meV (marked by the red triangle in Figs.~\ref{fig:BS_DOS_Po_shift_NonUniZGNR}(d) and~\ref{fig:BS_DOS_Po_shift_NonUniZGNR}(f)) demonstrates one of the characteristics of bilayer ZGNRs.
The DOS varies drastically when the top layer moves above the center of the bottom layer.
The QLL peaks are strongly suppressed, and most of them are replaced by lower-intensity asymmetric peaks corresponding to the band-edge states of the irregular oscillatory QLLs (Fig.~\ref{fig:BS_DOS_Po_shift_NonUniZGNR}(e)).
The width of the center-positioned top layer determines whether the DOS is closer to the monolayer- or bilayer-like DOS.
Monolayer-like DOS with additional asymmetric subpeaks are revealed for a very narrow top layer (Fig.~\ref{fig:BS_DOS_Width_var_NonUniZGNR}(d)).
When the top layer becomes sufficiently wide ($W_t > l_B$ in Fig.~\ref{fig:BS_DOS_Width_var_NonUniZGNR}(e)), the frequency spacings between the suppressed QLLs are reduced (red arrows), and the second group of QLLs forms at $\pm \gamma_1$ (blue arrows).
At $W_t = W_b$, the DOS of the bilayer ZGNR is displayed (Fig.~\ref{fig:BS_DOS_Width_var_NonUniZGNR}(f)).
The monolayer-like QLL peaks can concurrently exist with the bilayer-like QLL peaks for nonuniform ZGNRs with $W_t > l_B$ and $W_b - W_t > l_B$ as shown in Fig.~\ref{fig:BS_DOS_Coexistence_NonUniZGNR}(c) (in Fig.~\ref{fig:BS_DOS_Coexistence_NonUniZGNR}(d), $(W_b - W_t)/2 > l_B$).
The monolayer-like QLL peaks with wider spacings are indicated by the red circles, and the bilayer-like QLL peaks with narrower spacings are marked by the green triangles and blue diamonds.
Their relative peak height is sensitive to the top-layer position; for example, the intensity of the monolayer-like QLL peaks decreases rapidly for the case of center-positioned top layer (Fig.~\ref{fig:BS_DOS_Coexistence_NonUniZGNR}(d)).
It is predicted that the optical absorption peaks are contributed by transitions between high-DOS QLLs.

\begin{figure*}
\begin{center}
  \includegraphics[width=\linewidth, keepaspectratio]{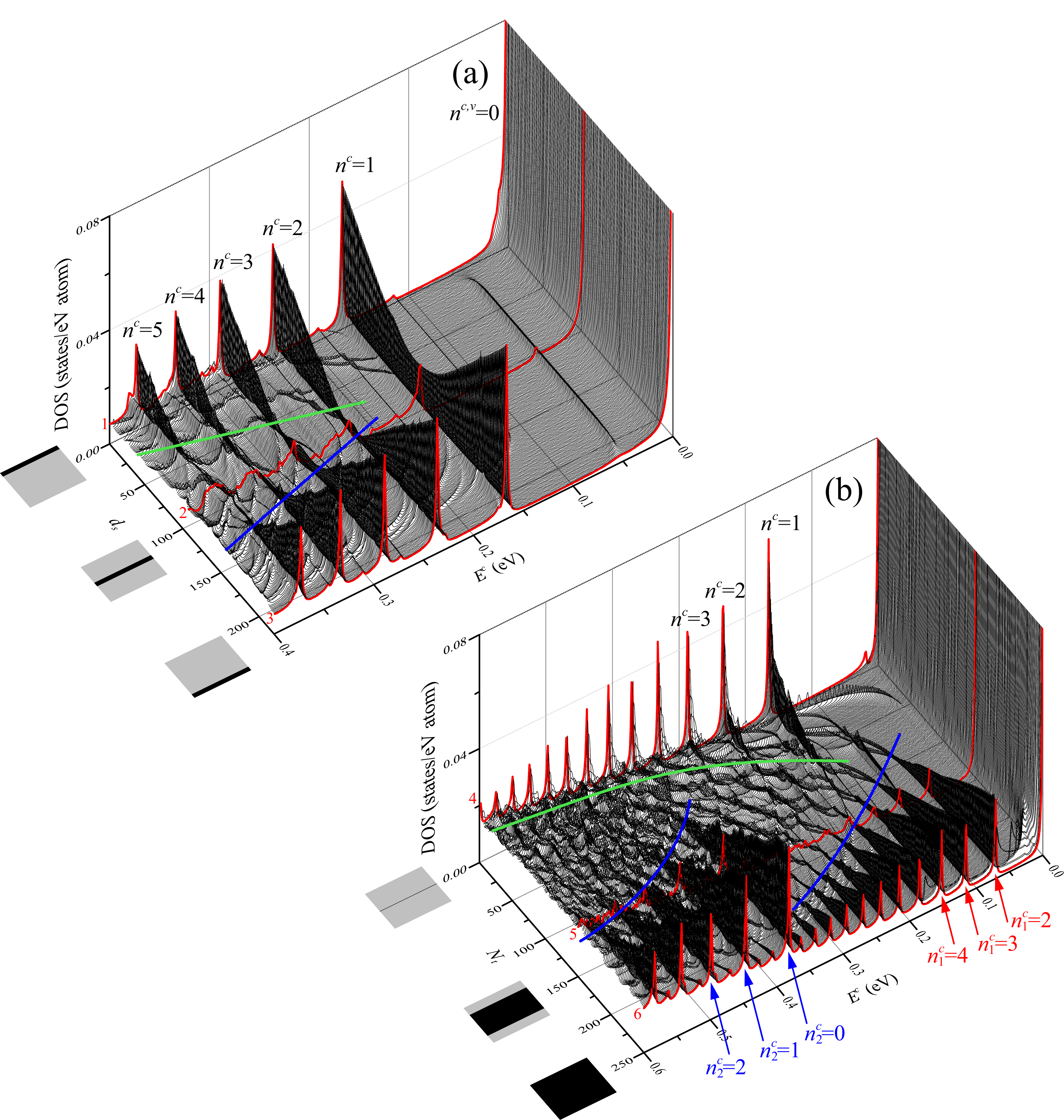}\\
  \caption[]{
  (a) Position-dependent DOS for the nonuniform ZGNR with $N_b = 250$ and $N_t = 25$.
  (b) Width-dependent DOS for the nonuniform ZGNR with $N_b = 250$ and a centered top layer.
  The DOSs at three characteristic positions (curves 1--3 for $d_s = 0, 112$, and $225$) and for three characteristic widths (curves 4--6 for $N_t = 2,150$, and $250$) are shown in red in Figs.~\ref{fig:BS_DOS_Po_shift_NonUniZGNR}(d)--\ref{fig:BS_DOS_Po_shift_NonUniZGNR}(f) and Figs.~\ref{fig:BS_DOS_Width_var_NonUniZGNR}(d)--\ref{fig:BS_DOS_Width_var_NonUniZGNR}(f).
  The geometric configurations for the characteristic positions/widths are indicated by the shaded parallelograms.
  }
  \label{fig:DOS_dynamic_variation_NonUniZGNR}
\end{center}
\end{figure*}

The position- and width-dependent DOSs can provide full information on the electronic energy spectra.
Special structures are formed or destroyed when the top-layer position changes ($0 \leq d_s \leq 225$ in Fig.~\ref{fig:DOS_dynamic_variation_NonUniZGNR}(a)).
The highest peak is contributed by the $n^{c,v} = 0$ QLL and degenerate partial flat subbands.
The intensity of the $n^c = 1$ QLL peak gradually decreases, reaches the minimum ($d_s = 112$ indicated by the red curve), and then increases to the maximum ($d_s = 225$).
This is due to the change from the dispersionless QLL to the irregular oscillatory QLL.
The other peaks disappear and form at certain critical $d_s$'s values (indicated by the green and blue lines).
This is relatively easily observed for the larger-$n^c$ QLL peaks since the dispersionless QLLs are thoroughly transformed into irregular oscillatory QLLs.
For another structure in the DOS, the ripples originate from many asymmetric peaks pertaining to the band-edge states of the irregular oscillatory QLLs.
On the other hand, the DOS is drastically altered as the top-layer width varies in the range of $2 \leq N_t \leq 250$ (Fig.~\ref{fig:DOS_dynamic_variation_NonUniZGNR}(b)).
The intensities of the monolayer-like QLL peaks decrease rapidly and reduce to zero at certain critical widths (green curve).
Two groups of QLL peaks with smaller energy spacings are revealed as the top-layer width increases (blue curves).
The unusual evolution of DOS reflects the drastic transformation from a monolayer-like QLL spectrum to a bilayer-like QLL spectrum.
Moreover, the distribution of the monolayer and bilayer electronic states are clearly distinguished, facilitating the determination of the  geometric-structure-dependent QLL optical transitions.



The main characteristics and the interlayer-atomic-interaction-induced evolution of prominent Landau peaks in the DOS can be verified using scanning tunneling spectroscopy (STS), which is an extension of STM~\cite{Phys.Rev.Lett.49(1982)57G.Binnig} and provides full information about the surface DOS, such as the DOS of silicon~\cite{Phys.Rev.Lett.56(1986)1972R.J.Hamers, Phys.Rev.B48(1993)17892M.N.Piancastelli} and CNTs~\cite{Nature391(1998)59J.W.G.Wildoer, Science283(1999)52L.C.Venema}.
The tunneling conductance ($dI/dV$) is proportional to the DOS, and therefore, the structures, positions, and intensities of peaks can be resolved.
In very wide GNRs, low-frequency symmetric Landau peaks showing  $\sqrt{n^{c,v}B_0}$ dependence have been observed~\cite{Science324(2009)924D.L.Miller, Nature467(2010)185Y.J.Song}.
The aforementioned four categories of magnetoelectronic energy spectra in  nonuniform ZGNRs can be identified through STS measurements of the DOS~\cite{Phys.Rev.B71(2005)193406Y.Kobayashi, Phys.Rev.Lett.97(2006)236804Y.Niimi, Science324(2009)924D.L.Miller}.
Moreover, the variations of the prominent symmetric QLL peaks during a change in the top layer position/width can be observed, including the peak strength, formation/disapperance of peaks, and the transformation from monolayer-like QLLs to bilayer-like QLLs.
In particular, the coexistence of monolayer- and bilayer-like QLLs has been evidenced by conductance measurements.
Through gate voltage adjustment, strong suppression of the bilayer resistance oscillation has been observed~\cite{Phys.Rev.B79(2009)235415C.P.Puls}.

Wave functions with diverse spatial distributions are crucial for realizing fundamental physical properties, such as charge density~\cite{Phys.Rev.28(1926)1049E.Schrodinger, Phys.Rev.B78(2008)235311M.W.Y.Tu, Phys.Rev.B89(2014)121401P.M.PerezPiskunow, Phys.Rev.B90(2014)115423G.Usaj}, state mixing~\cite{J.Phys.Soc.Jpn.80(2011)044602H.C.Chung}, and optical transition~\cite{NanoLett.6(2006)2748V.Barone, J.Appl.Phys.103(2008)073709Y.C.Huang}.
The wave functions of ZGNRs (Eq. (\ref{eq:WFOfNonUniGNRs})) can be decomposed into the subenvelope functions of distinct sublattices as follows:
\begin{eqnarray}
\nonumber
| \Psi^{c,v} \rangle & = &
 \displaystyle\sum_{m=1,3,5,...}^{}
\bigg( A^1_o |\Phi^{A^1}_m\rangle + B^1_o |\Phi^{B^1}_m\rangle
     + A^2_o |\Phi^{A^2}_m\rangle + B^2_o |\Phi^{B^2}_m\rangle \bigg) \\
& + & \displaystyle\sum_{n=2,4,6,...}^{}
\bigg( A^1_e |\Phi^{A^1}_n\rangle + B^1_e |\Phi^{B^1}_n\rangle
     + A^2_e |\Phi^{A^2}_n\rangle + B^2_e |\Phi^{B^2}_n\rangle \bigg) ,
\label{eq:WFDecomposedOfNonUniGNRs}
\end{eqnarray}
where the subscript $o$ ($e$) indicates amplitudes on odd (even) zigzag lines. Only the odd terms are discussed because of the phase difference of $\pi$ (e.g., $A^1_o = - A^1_e$).
In general, the electronic states in the parabolic subbands present standing waves, reflecting the lateral confinement.
Furthermore, the quantized states in the dispersionless QLLs show Landau wave functions, which are products of a Hermite polynomial and a Gaussian distribution function.
Because the prominent absorption peaks arise from the optical transitions between QLLs, it is worthwhile to investigate the wave functions at the formation center of QLLs.

\begin{figure}
\begin{center}
  \includegraphics[width=\linewidth, keepaspectratio]{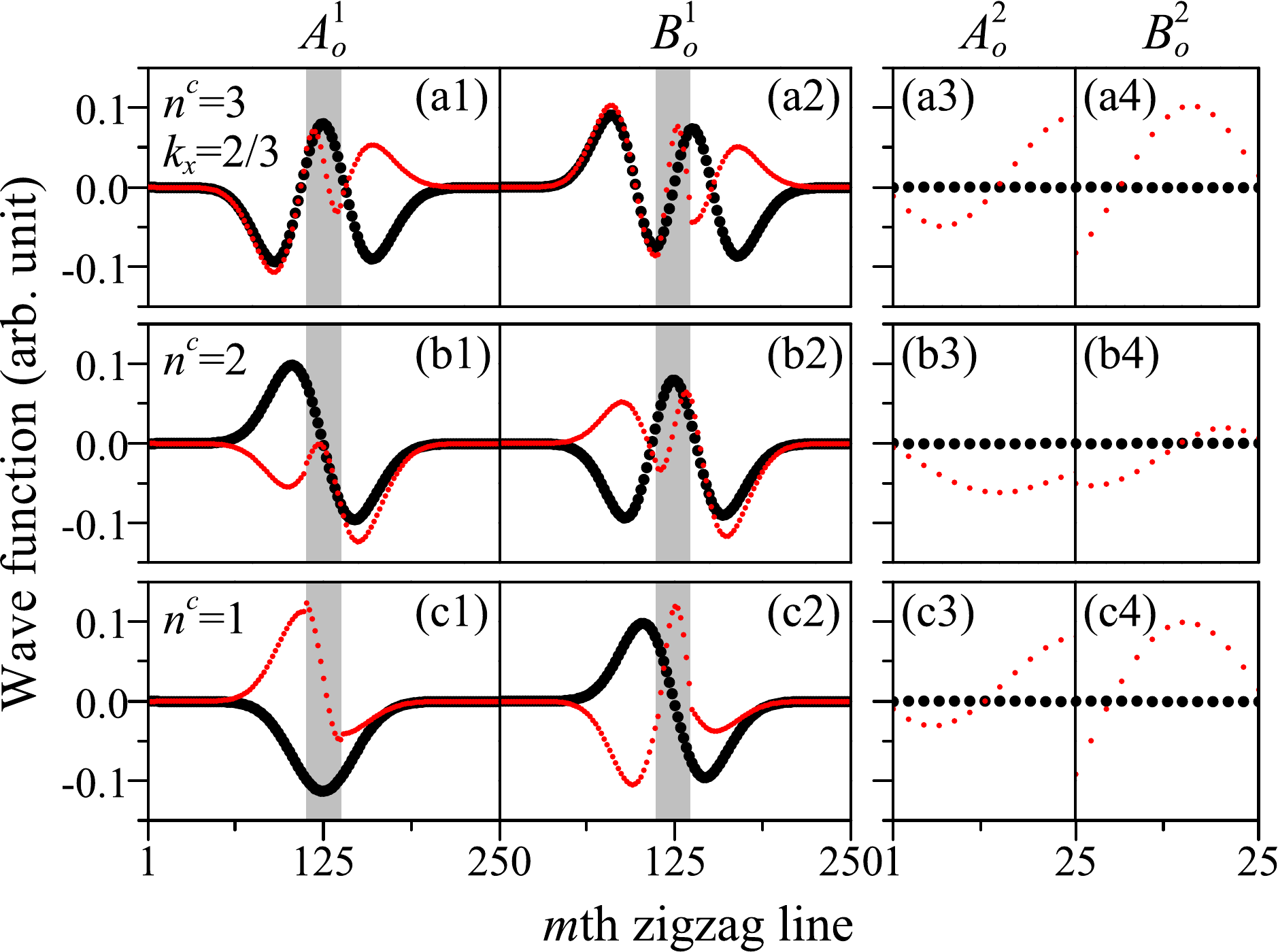}\\
  \caption[]{
  Wave functions of the subbands corresponding to the $n^c$ range $1$--$3$ at $k_x = 2/3$.
  Heavy black dots denote the top layer of width $N_t = 25$ at $d_s = 0$, and light red dots correspond to $N_t = 25$ and $d_s = 112$.
  The gray zones indicate the centered top layer.
  }
  \label{fig:WF_Po_shift_NonUniZGNR}
\end{center}
\end{figure}

The wave functions are strongly dependent on the position of the narrow top layer.
When the top layer is positioned at the edge of the bottom layer, the QLLs remain dispersionless and the center-located Landau wave functions are almost unaffected.
The $n^c = 1$--$3$ Landau wave functions at $k_x = 2/3$ (black circles in Fig.~\ref{fig:BS_DOS_Po_shift_NonUniZGNR}(a)) show Landau modes in $A_o^1$ and $B_o^1$ sublattices, and these modes are similar to those of the monolayer ZGNR (heavy black dots in Fig.~\ref{fig:WF_Po_shift_NonUniZGNR}).
Such spatial distributions of Landau modes are either symmetric or antisymmetric, and the node number of the subenvelope functions in sublattice $A_o^1$ is lower than that in sublattice $B_o^1$ by one.
For instance, there are one and two nodes on the $A_o^1$ and $B_o^1$ sublattices for the $n^c = 2$ QLL, respectively (Figs.~\ref{fig:WF_Po_shift_NonUniZGNR}(b1) and~\ref{fig:WF_Po_shift_NonUniZGNR}(b2)).
Therefore, the subband of the larger state energy $|E^{c,v}|$ has a larger node number, and the number of nodes on the $B_o^1$ sublattice is chosen to be suitable for indexing the subbands.
The electronic states are drastically altered by interlayer atomic interactions.
The states of the $n^c = 1$--$3$ irregular oscillatory QLLs at $k_x = 2/3$ (red circles in Fig.~\ref{fig:BS_DOS_Po_shift_NonUniZGNR}(b)) are mixed with the states of parabolic subbands, and therefore, the wave functions (light red dots in Fig.~\ref{fig:WF_Po_shift_NonUniZGNR}) become piecewise continuous or have two jump discontinuities in the $A_o^1$ and $B_o^1$ sublattices at the two edges of the top ribbon (edges of gray zone), indicating the absence of the normal Landau modes.
Furthermore, the spatial distributions of the top layer ($A_o^2$ and $B_o^2$) turn out to be observable, showing distorted standing waves.
It is predicted that the suppression of QLL transition peaks and the appearance of additional subpeaks occurs because of the distortions in wave functions.

\begin{figure}
\begin{center}
  \includegraphics[width=\linewidth, keepaspectratio]{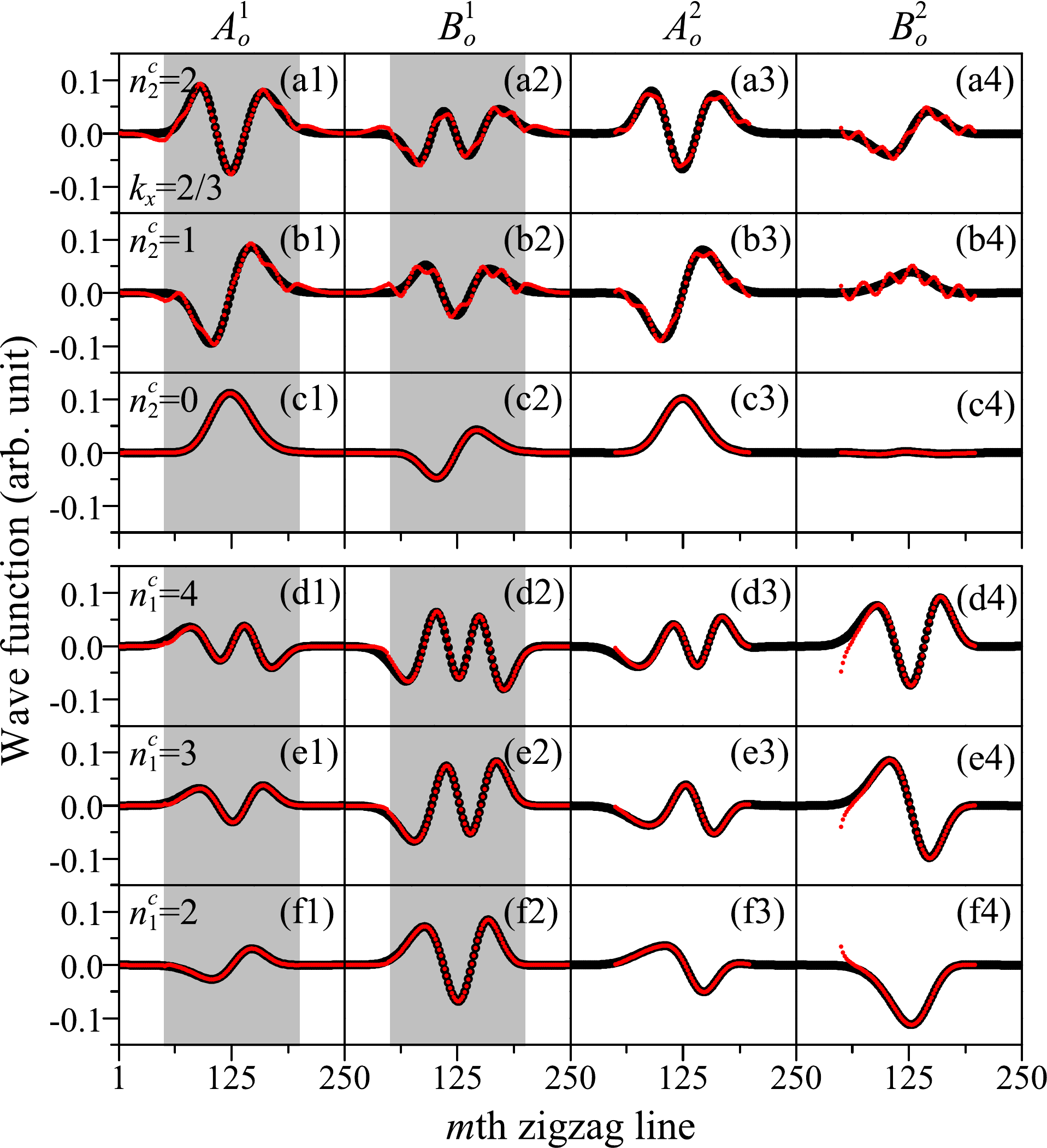}\\
  \caption[]{
  Wave functions of the low-lying QLLs at $k_x = 2/3$.
  The light red dots denote the centered top layer of width $N_t = 150$, and the heavy black dots represent the centered top layer of width $N_t = 250$.
  The gray zones indicate the centered top layer with $N_t = 150$.
  The locations of the electronic states on the first- and second-group QLLs are marked by the red triangles and blue diamonds in Fig.~\ref{fig:BS_DOS_Width_var_NonUniZGNR}, respectively.
  }
  \label{fig:WF_Width_var_NonUniZGNR}
\end{center}
\end{figure}

The monolayer-like Landau wave functions evolve into bilayer-like Landau wave functions when the top layer of the nonuniform GNR becomes sufficiently wide.
In the uniform bilayer ZGNRs ($W_b = W_t$), the spatial distributions of Landau wave functions, including the amplitude, spatial symmetry, and node number, are not well behaved under the influence of interlayer interactions.
In both groups of Landau states, the wave functions at the formation center are somewhat distorted and the amplitudes are different for the four sublattices (heavy black dots in Fig.~\ref{fig:WF_Width_var_NonUniZGNR}).
Obviously, in the first-group (second-group) Landau states, the amplitude of the $B_o^1$ ($A_o^1$) sublattice dominates among the amplitudes of the four sublattices, and the corresponding node number is characterized as the subband index.
In other words, for the $n_1^{c,v} = n \geq 2$ ($n_2^{c,v} = n \geq 1$) Landau wave functions, there are $n-1$, $n$, $n-1$, and $n-2$ ($n$, $n+1$, $n$, and $n-1$) nodes in the subenvelope functions of sublattices $A_o^1$, $B_o^1$, $A_o^2$, and $B_o^2$, respectively.
For the nonuniform ZGNR with a sufficiently wide top layer ($W_b > W_t > l_B$), the QLL wave functions are similar to those of bilayer ZGNRs (light red dots in Fig.~\ref{fig:WF_Width_var_NonUniZGNR}).
The regularity of node numbers in the Landau modes and the dominant sublattices for each group remain the same.
However, the Landau wave functions of the second group have an extra component---the standing waves of parabolic subbands---and therefore show slightly distorted distributions (Figs.~\ref{fig:WF_Width_var_NonUniZGNR}(a) and~\ref{fig:WF_Width_var_NonUniZGNR}(b)).
It is predicted that there are four categories of inter-QLL transitions, two intragroup and two intergroup transitions.

\begin{figure}
\begin{center}
  \includegraphics[width=\linewidth, keepaspectratio]{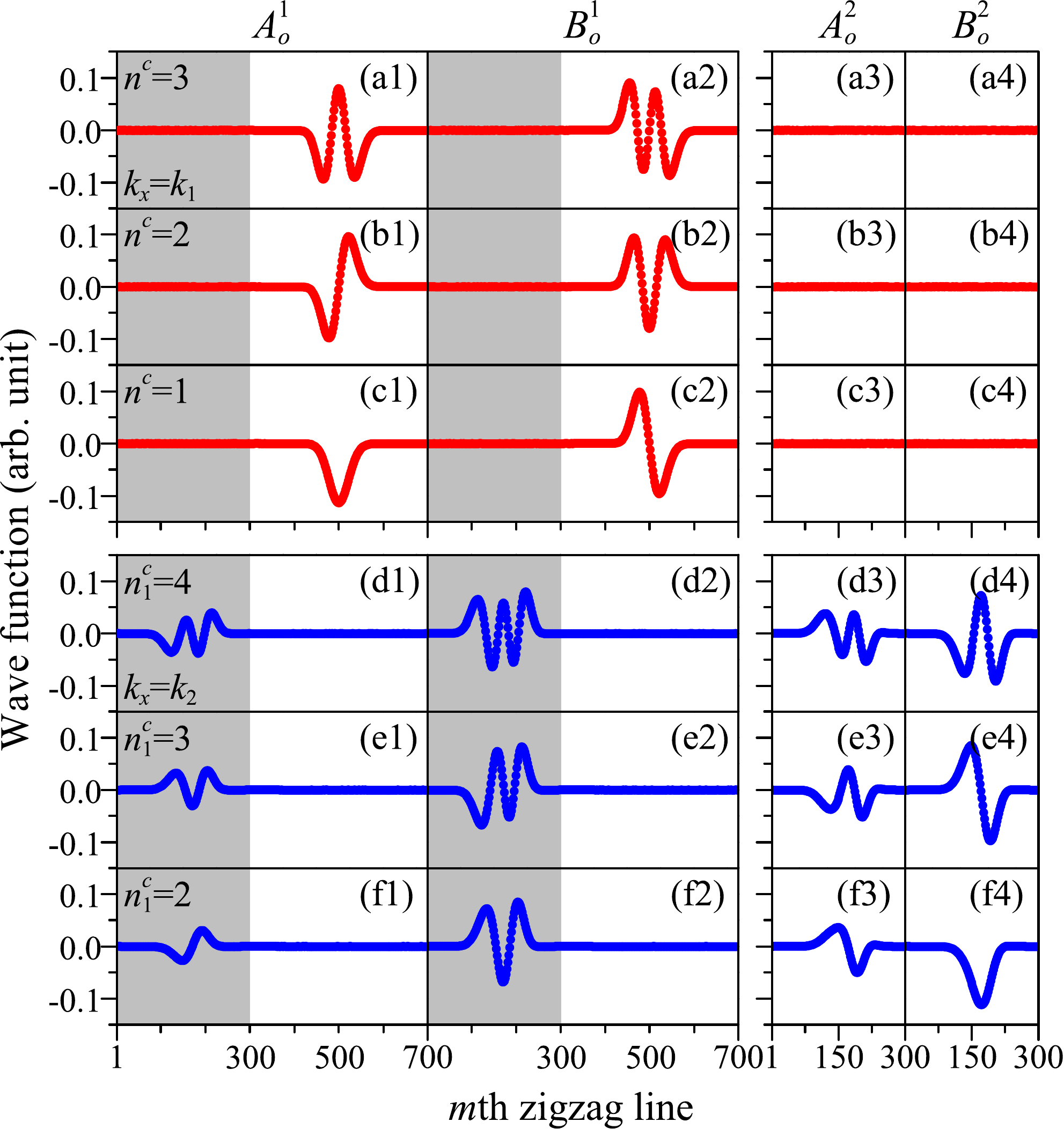}\\
  \caption[]{
  Wave functions of the low-lying subbands at $k_x = k_1$ and $k_2$.
  The shaded regions represent the top layer of width $N_t = 300$ at $d_s = 0$.
  The locations of the electronic states at $k_x = k_1$ and $k_2$ are indicated by the red and blue crosses in Fig.~\ref{fig:BS_DOS_Coexistence_NonUniZGNR}(a), respectively.
  }
  \label{fig:WF_Coexistence_NonUniZGNR}
\end{center}
\end{figure}

For the nonuniform GNR with sufficiently wide overlapping and nonoverlapping regions, the monolayer and bilayer Landau wave functions can concurrently exist.
At the formation center of monolayer-like QLLs ($k_x = k_1$), the features of Landau wave functions are identical to those of the QLLs in monolayer ZGNR.
The symmetric/antisymmetric subenvelope functions of sublattices $A_o^1$ and $B_o^1$ are localized at the center of the nonoverlapping region (white zones in Figs.~\ref{fig:WF_Coexistence_NonUniZGNR}(a1),
~\ref{fig:WF_Coexistence_NonUniZGNR}(a2),
~\ref{fig:WF_Coexistence_NonUniZGNR}(b1),
~\ref{fig:WF_Coexistence_NonUniZGNR}(b2),
~\ref{fig:WF_Coexistence_NonUniZGNR}(c1),
and~\ref{fig:WF_Coexistence_NonUniZGNR}(c2)), and their node numbers are $n-1$ and $n$ for the $n^{c,v} = n$ monolayer-like QLLs, respectively.
Meanwhile, at the formation center of bilayer-like QLLs ($k_x = k_2$), Landau wave functions of the bilayer ZGNR are exhibited.
The slightly distorted subenvelope functions are localized at the center of the overlapping region (Figs.~\ref{fig:WF_Coexistence_NonUniZGNR}(d)--\ref{fig:WF_Coexistence_NonUniZGNR}(f)), and the node number in each sublattice is identical to that of the bilayer ZGNR.
In addition, near the interface between the two regions, the wave functions are distributed in a piecewise continuous form and the distorted bilayer and monolayer Landau wave functions are joined~\cite{Phys.Rev.B82(2010)205436M.Koshino}.
The prominent absorption spectrum is predicted to be characterized by optical transitions between the monolayer-like QLLs and between the bilayer-like QLLs.


The aforementioned features of wave functions can be investigated by spectroscopic-imaging STM~\cite{Science262(1993)218M.F.Crommie, Nature427(2004)328S.Ilani, Science327(2010)665A.Richardella}.
It can resolve charge distributions from the local DOS and is a powerful tool for identifying standing waves and Landau wave functions on the surfaces of various condensed-matter systems.
Standing waves have been directly observed at the surface steps of Au(111) and Cu(111)~\cite{Nature363(1993)524M.F.Crommie, Science262(1993)218M.F.Crommie}.
Furthermore, theoretical predictions of the standing waves in a finite-length metallic CNT~\cite{Phys.Rev.Lett.82(1999)3520A.Rubio, Phys.Rev.B65(2002)245418J.Jiang} have also been verified~\cite{Science283(1999)52L.C.Venema, Phys.Rev.Lett.93(2004)166403J.Lee}.
Recently, Landau orbits without the zero point have been observed~\cite{Phys.Rev.Lett.101(2008)256802K.Hashimoto, Nat.Phys.6(2010)811D.L.Miller}, and subsequently, observations of the concentric-ring-like nodal structures have also been made~\cite{Phys.Rev.Lett.109(2012)116805K.Hashimoto, Nat.Phys.10(2014)815Y.S.Fu}.
In nonuniform GNRs, the monolayer/bilayer/distorted Landau wave functions are closely related to the geometric configuration.
In particular, interlayer interactions lead to state mixing, and interfaces give rise to  the discontinuous distributions.
The diverse spatial distributions in Landau wave functions can be examined through spectroscopic-imaging STM measurements on nodal structures.


\section{Magneto-optical properties}

When a GNR is placed in an electromagnetic field with the electric polarization $\hat{\mathbf{e}} \Vert \hat{x}$, electrons are excited from occupied to unoccupied states.
Vertical optical excitations with the same wave vector ($\Delta k_x = 0$) are available since the photon momentum is almost zero.
The optical absorption function based on Fermi's golden rule is given by
\begin{eqnarray}
\nonumber
A( \omega ) & \propto &
 \displaystyle\sum_{h,h'=c,v} \sum_{n,n'} \int_{\mathrm{B.Z.}} \frac{dk_{x}}{2\pi}
 \mathrm{Im} \bigg\{ \frac{f[E^{h'}(k_x,n')]-f[E^{h}(k_x,n)]}{E^{h'}(k_x,n')-E^{h}(k_x,n)-\omega-i\Gamma} \bigg\} \\
\displaystyle & \times &
\left| \bigg\langle\Psi^{h'}(k_x,n')\bigg|\frac{\hat{\mathbf{e}}\cdot\mathbf{p}}{m_e}\bigg|\Psi^{h}(k_x,n)\bigg\rangle \right| ^2 ,
\label{eq:AbsFormulaForGNR}
\end{eqnarray}
where $m_{e}$ is the bare electron mass, $\mathbf{p}$ is the momentum operator, $f[E^h(k_x,n)]$ is the Fermi--Dirac distribution function, and $\Gamma$ ($\approx 2$ meV) is the phenomenological broadening parameter.
The spectral intensity is dominated by the joint DOS (the first term) and the velocity matrix element (the second term).
The second term is evaluated within the gradient approximation~\cite{Phys.Rev.160(1967)649G.Dresselhaus, Phys.Rev.B7(1973)2275L.G.Johnson, Phys.Rev.B47(1993)15500L.C.LewYanVoon}; such evaluation has previously been successfully performed for layered graphite~\cite{Phys.Rev.B7(1973)2275L.G.Johnson, Phys.Rev.B67(2003)165402A.Gruneis}, CNTs~\cite{Phys.Rev.B67(2003)165402A.Gruneis, Carbon42(2004)3169J.Jiang}, and few-layer graphenes~\cite{NewJ.Phys.12(2010)083060C.W.Chiu, ACSNano4(2010)1465Y.H.Ho}.
The velocity matrix element is expressed as
$(1/\hbar ) \sum_{{l,l'}=1} C_l^{h'*}(k_x,n') C_{l'}^h(k_x,n) [\partial H_{l,l'}(k_x) / \partial k_x]$,
where $C_{l'}^h(k_x,n)$ is the amplitude at the lattice site of the subenvelope  function.
The nonvanishing $\partial H_{l,l'}/\partial k_x$ is related to the nearest-neighbor atomic interactions.
The presence of optical transitions depends on the finite inner product among the initial state and final state on different sublattices and $\partial H_{l,l'}/\partial k_x$.
For example, the edge-dependent optical selection rules for GNRs in the absence of external fields have been obtained by analyzing the velocity matrix elements and the special relationships among wave functions.
The selection rules are $|\Delta J| = odd$ for ZGNRs and $\Delta J = 0$ for AGNRs, where $J$ is the subband index~\cite{Phys.Rev.B76(2007)045418H.Hsu, Opt.Express19(2011)23350H.C.Chung}.

\begin{figure}
\begin{center}
  \includegraphics[width=\linewidth, keepaspectratio]{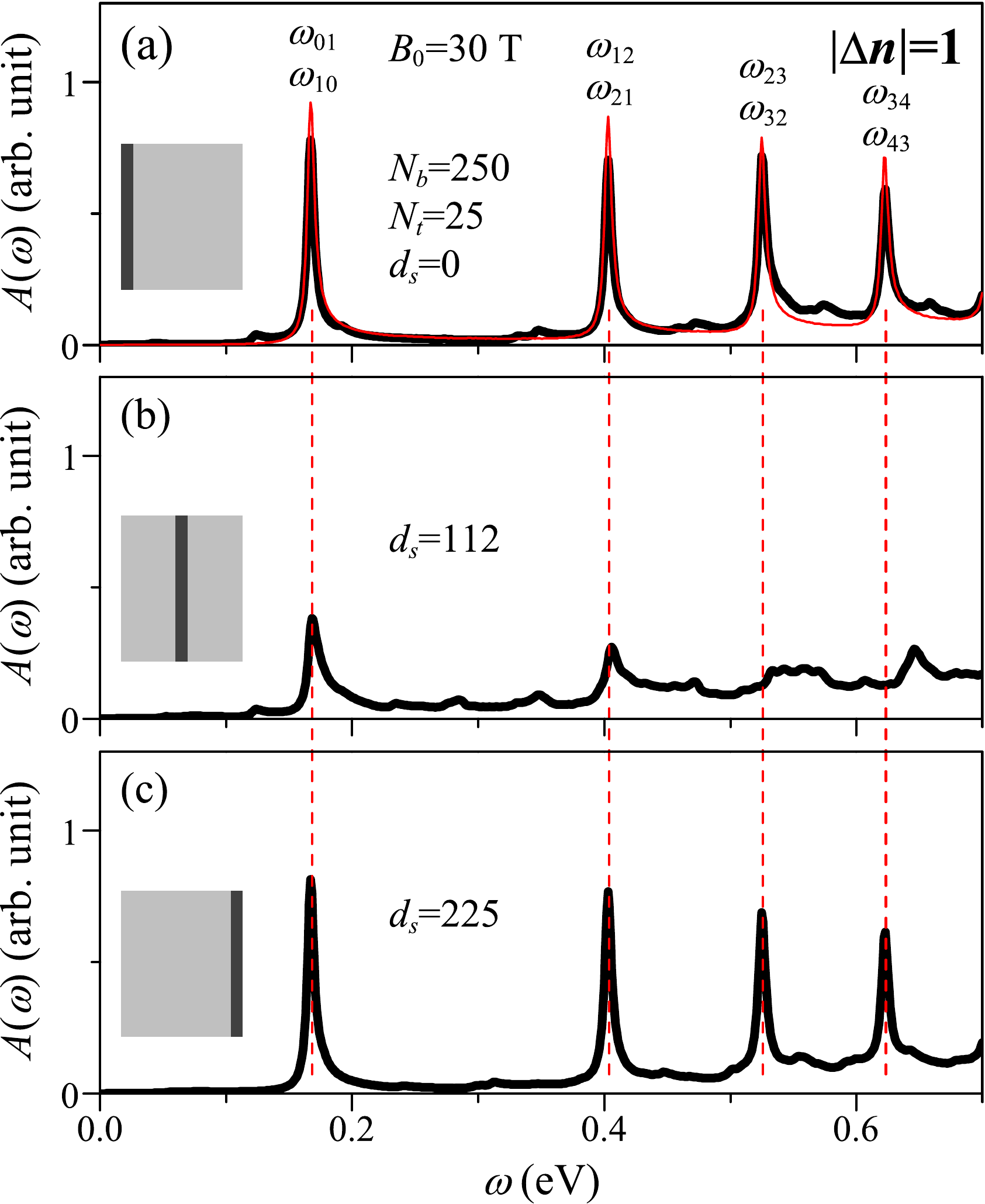}\\
  \caption[]{
  Low-frequency magneto-optical absorption spectra of the nonuniform ZGNRs with $N_b = 250$ and $N_t = 25$ at (a) $d_s =0$, (b) $d_s = 112$, and (c) $d_s = 225$ for $B_0 = 30$ T.
  Light red curve indicates the magnetoabsorption spectrum of the monolayer ZGNR.
  Dashed red lines indicate the inter-QLL absorption frequencies of the monolayer ZGNR.
  }
  \label{fig:ABS_Po_shift_NonUniZGNR}
\end{center}
\end{figure}

On the basis of the geometric configuration, the magnetoabsorption spectra of  nonuniform GNRs can be divided into four categories: monolayer-like, bilayer-like, coexistent, and irregular spectra.
The prominent QLL transition peaks might be strongly suppressed, depending on the position of the narrow top layer.
For an edge-aligned top layer (Figs.~\ref{fig:ABS_Po_shift_NonUniZGNR}(a) and~\ref{fig:ABS_Po_shift_NonUniZGNR}(c)), the magnetoabsorption spectrum is very similar to that of monolayer ZGNRs (light red curve), except for few extra subpeaks.
There are many prominent symmetric peaks ($\omega_{n^v n^c}$'s) associated with the vertical transitions from the $n^v$ QLL to the $n^c = n^v \pm 1$ QLL.
This is because valence and conduction subbands with adjacent indices possess the same Landau mode.
The velocity matrix elements have finite values, and therefore, the optical transitions can exist for the subbands satisfying the magneto-optical selection rule $|\Delta n| = | n^c - n^v | =  1$~\cite{Nanotechnol.18(2007)495401Y.C.Huang, J.Appl.Phys.103(2008)073709Y.C.Huang}.
Furthermore, the peak height, which is proportional to the $k_x$ range of the QLL, decreases with an increase in the frequency.
Notably, the QLL transition frequencies exhibit the simple relation $\sqrt{B_0}(\sqrt{n^{v}} + \sqrt{n^{v} \pm 1})$ based on the $\sqrt{n^{c,v}B_0}$ relationship and the selection rule $|\Delta n| = 1$.
The extra absorption subpeaks originate from the transitions between the band-edge states of mixed subbands.
In particular, the number, intensity, and frequency of the absorption peaks are very sensitive to the position of the top layer, especially for a center-positioned top layer (Fig.~\ref{fig:ABS_Po_shift_NonUniZGNR}(b)).
Most of the higher-frequency peaks are thoroughly suppressed because of  distortions in the wave functions, whereas the intensity of the low-frequency peaks is reduced.

\begin{figure}
\begin{center}
  \includegraphics[width=0.87\linewidth, keepaspectratio]{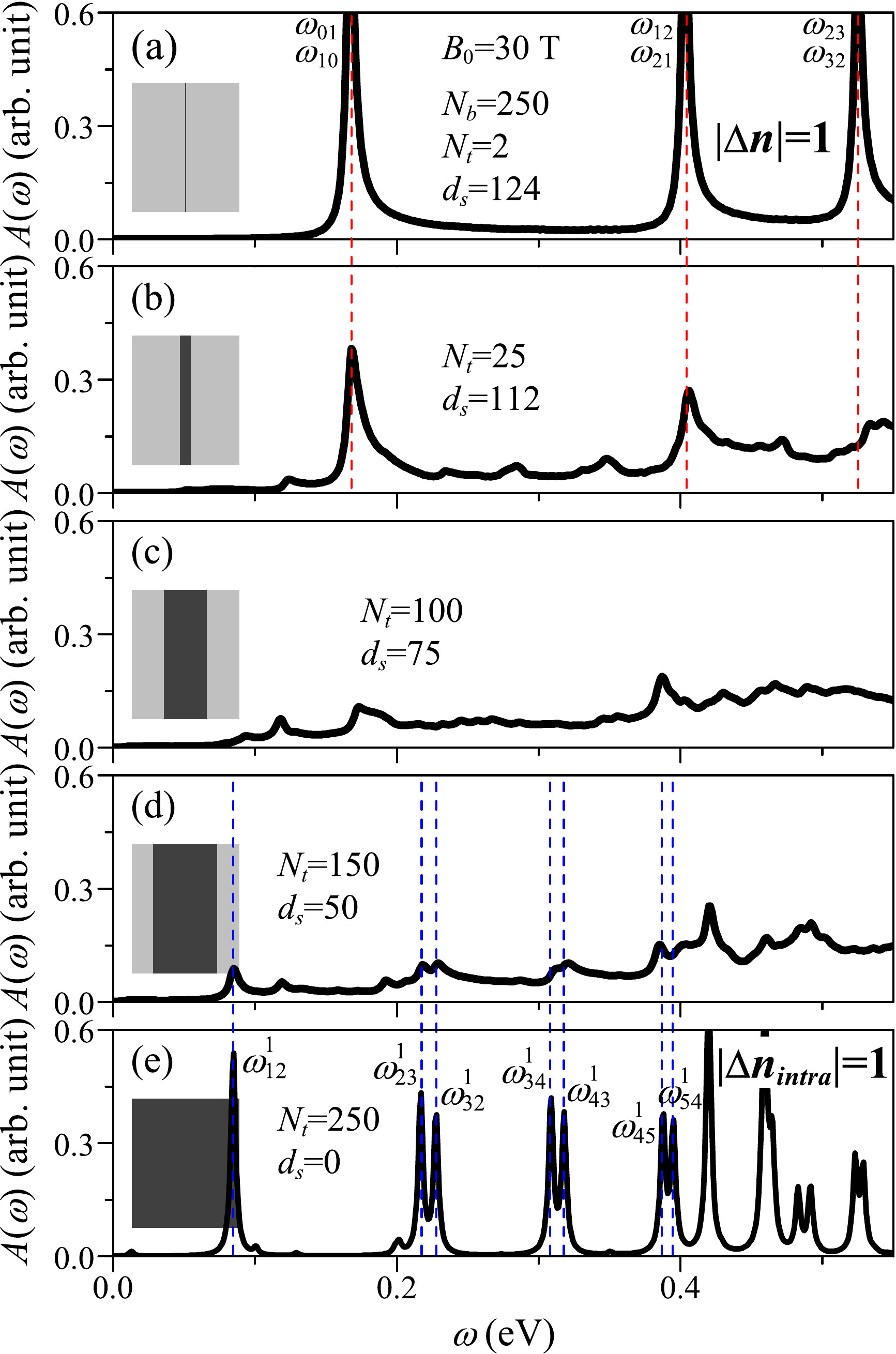}\\
  \caption[]{
  Magneto-optical absorption spectra of the nonuniform ZGNRs with $N_b = 250$ for (a) $N_t = 2$, (b) $N_t = 25$, (c) $N_t = 100$, (d) $N_t = 150$, and (e) $N_t = 250$ at the ribbon center.
  Dashed red and blue lines represent the inter-QLL absorption frequencies of the monolayer and bilayer ZGNRs, respectively.
  }
  \label{fig:ABS_Width_var_NonUniZGNR}
\end{center}
\end{figure}

The top-layer width dominates whether the QLL spectrum is a monolayer-like spectrum or bilayer-like spectrum.
Monolayer-like QLLs are exhibited in the nonuniform GNR with a very narrow top layer (Fig.~\ref{fig:ABS_Width_var_NonUniZGNR}(a)).
The intensity of the QLL transition peaks is lower for a wider top layer (Fig.~\ref{fig:ABS_Width_var_NonUniZGNR}(b)).
When the top ribbon width is close to the spatial distribution of Landau wave functions ($N_t = 100$ in Fig.~\ref{fig:ABS_Width_var_NonUniZGNR}(c)), the interlayer interactions strongly suppress the QLL transition peaks, lead to the absence of selection rules, and cause many low-intensity subpeaks to appear.
The main reason is that the dispersionless QLLs become irregular oscillatory QLLs, and the wave functions are highly distorted.
However, two groups of QLLs are formed as the top layer width is sufficient for magnetic quantization (e.g., $N_t = 150$ in Fig.~\ref{fig:BS_DOS_Width_var_NonUniZGNR}(b)).
Consequently, the absorption spectrum ($N_t = 150$ in Fig.~\ref{fig:ABS_Width_var_NonUniZGNR}(d)) is similar to that of the bilayer  ($W_t = W_b$ in Fig.~\ref{fig:ABS_Width_var_NonUniZGNR}(e)).
There are four categories of inter-QLL absorption peaks, and half of them belong to intragroup or intergroup excitations.
In particular, the low-frequency peaks originate from the intragroup QLL transitions ($\omega_{{n_1^c}{n_1^v}}^1$) between the $n_1^c$ and $n_1^v$ QLLs.
These peaks obey the selection rule $| \Delta n_{intra} | = | n_1^c - n_1^v | = 1$ because the Landau modes are the same on the $i$th layer for the $A^i$ ($B^i$) sublattice of the initial state and the $B^i$ ($A^i$) sublattice of the final state~\cite{Phys.Rev.B78(2008)115422Y.C.Huang, Phys.Chem.Chem.Phys.18(2016)7573H.C.Chung}.
Because of the asymmetric QLL spectrum, the intragroup transitions ($\omega_{n(n-1)}^1$ and $\omega_{(n-1)n}^1$) show twin-peak structures,   except for the threshold transitions ($\omega_{12}^1$).

\begin{figure}
\begin{center}
  \includegraphics[width=\linewidth, keepaspectratio]{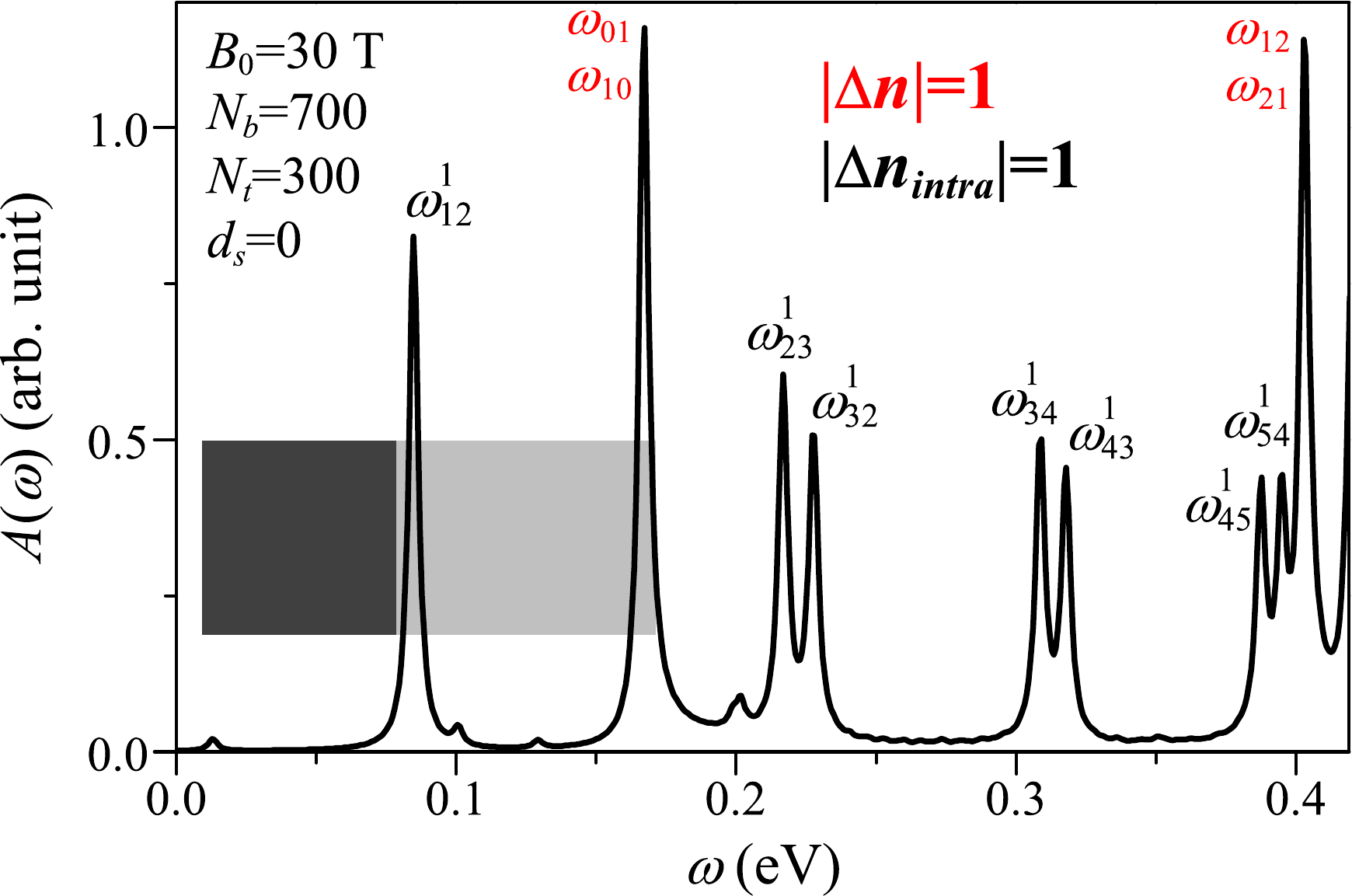}\\
  \caption[]{
  Magneto-optical spectrum of the nonuniform ZGNR with $N_b = 700$ and $N_t = 300$ at $d_s = 0$.
  }
  \label{fig:ABS_Coexistence_NonUniZGNR}
\end{center}
\end{figure}

Monolayer and bilayer QLL transitions can exist simultaneously when the nonoverlapping and overlapping regions are sufficiently wide for the quantization of the electronic states.
The well-behaved monolayer and bilayer Landau wave functions are in the nonoverlapping and overlapping regions, respectively (Fig.~\ref{fig:WF_Coexistence_NonUniZGNR}), and therefore, the different optical transitions can concurrently exist.
The $|\Delta n| = 1$ absorption peaks with wide frequency spacings and the $|\Delta n_{intra}| = 1$ peaks with narrower frequency spacings are clearly present in Fig.~\ref{fig:ABS_Coexistence_NonUniZGNR}.
Moreover, their relative peak height is roughly proportional to the ratio between the widths of the two distinct regions.

The inter-QLL transitions and selection rules can be examined through experimental measurements, as previously performed for few-layer graphene-related systems by using magneto-optical transmission~\cite{Phys.Rev.Lett.97(2006)266405M.L.Sadowski, Phys.Rev.Lett.110(2013)246803J.M.Poumirol} and magneto-Raman spectroscopy~\cite{NanoLett.14(2014)4548S.Berciaud}.
Some of the theoretical predictions, especially regarding the threshold absorption peak of $\omega_{01}$ in a magnetic-field, of GNRs have been recently verified through infrared transmission measurement on epitaxial monolayer GNR arrays~\cite{Phys.Rev.Lett.110(2013)246803J.M.Poumirol}.
For narrow monolayer GNRs, $\omega_{01}$ does not obey the $\sqrt{B_0}$ dependence at low magnetic fields.
However, the dependence is satisfied at high magnetic fields.
The predicted magneto-optical properties of nonuniform GNRs can be verified, and they include the monolayer, bilayer, and coexistent QLL absorption spectra, magneto-optical selection rules, and single-peak, twin-peak, and irregular-peak structures.
They are quite sensitive to the geometric configuration.
Moreover, the discussions presented provide an elementary example for understanding the optical properties of other similar structures and the effect of van der Waals interactions, such as partially overlapping GNRs~\cite{Science336(2012)1143A.W.Tsen, Carbon75(2014)411M.Berahman, Carbon80(2014)513R.Rao}, GNR-graphene superlattices~\cite{J.Phys.Chem.C117(2013)7326J.H.Wong}, and van der Waals heterostructures~\cite{Nature499(2013)419A.K.Geim, Nanoscale7(2015)13393Z.Zheng, Phys.Rev.B93(2016)075438L.E.F.FoaTorres}.

\section{Conclusion}

The low-energy magnetoelectronic and optical properties of nonuniform bilayer GNRs can be considerably altered by changing their geometric configuration.
There are four categories of QLL spectra: monolayer-like, bilayer-like, coexistent, and irregular oscillatory spectra.
Monolayer-like QLLs are revealed for an edge-aligned narrow top layer ($W_t < l_B$), for which the Landau wave functions have almost well-behaved modes.
The prominent symmetric DOS peaks arise from the QLLs and partial flat subbands.
With an increase in the width of the top layer, the QLLs evolve from dispersionless to irregular oscillatory and show many extra band-edge states.
Furthermore, interlayer atomic interactions induce considerable distortions in wave functions and severe state mixing between parabolic subbands and QLLs.
The major symmetric DOS peaks are suppressed, and additional asymmetric subpeaks related to the extra band-edge states are revealed.
A bilayer-like QLL spectrum is formed when the top layer is sufficiently wide ($W_t > l_B$) for the quantization of the electronic states.
There are two groups of QLLs and symmetric DOS peaks.
Furthermore, the Landau wave functions become bilayer-like wave functions because of interlayer interactions.
For nonuniform GNRs with sufficiently wide nonoverlapping and overlapping regions, monolayer- and bilayer-like QLLs concurrently exist; moreover, DOS peaks are also revealed.
The well-behaved monolayer and bilayer Landau states are present in the nonoverlapping and overlapping regions, respectively.
The main characteristics of electronic properties revealed in the DOS and wave functions can be identified by using STS and spectroscopic-imaging STM.

Unusual electronic structures are reflected in the absorption spectra.
The low-frequency prominent symmetric absorption peaks corresponding to the inter-QLL transitions obey the magneto-optical selection rule $|\Delta n| = 1$ for monolayer-like QLLs and $|\Delta n_{intra}| = 1$ for bilayer-like QLLs.
In the disordered QLL spectrum, higher-frequency inter-QLL transitions are suppressed, and many low-intensity peaks without selection rules are revealed.
In particular, monolayer- and bilayer-like absorption spectra can coexist when the widths of both nonoverlapping and overlapping regions are sufficiently wide for magnetic quantization.
The feature-rich absorption spectra and selection rules can be explored using magneto-optical transmission and magneto-Raman spectroscopy.
Recently, partially overlapping structures between CVD-grown graphene grains have been synthesized~\cite{ACSNano5(2011)6610A.W.Robertson, Science336(2012)1143A.W.Tsen, Carbon80(2014)513R.Rao}.
The theoretical study of nonuniform GNRs can provide an essential understanding of such structures and the effective van der Waals interactions have on their properties.

\section*{Acknowledgements}
We would like to thank all the contributors to this article for their valuable discussions and recommendations, especially Ming-Hsun Lee, Po-Shin Shi, Thi-Nga Do, Ngoc Thanh Thuy Tran, Matisse Wei-Yuan Tu, Ping-Yuan Lo, Yu-Ming Wang, Hao-Chun Huang, and Kuan-Yu Chen.
The authors thank Pei-Ju Chien for English discussions and corrections.
One of us (Hsien-Ching Chung) thanks Ming-Hui Chung and Su-Ming Chen for financial support.
This research received funding from the Headquarters of University Advancement at the National Cheng Kung University, which is sponsored by the Ministry of Education, Taiwan.
This work was supported in part by the National Science Council of Taiwan under grant number NSC 102-2112-M-006-007-MY3.

\section*{References}

\bibliography{Reference}

\end{document}